\documentclass[fleqn, letterpaper, 10 pt, conference]{ieeeconf}  

\IEEEoverridecommandlockouts                              

\let\proof\relax

\usepackage{epsfig} 
\usepackage{amsthm}
\usepackage{amsmath} 
\usepackage{amssymb}  

\usepackage{algorithm}
\usepackage{algpseudocode}

\usepackage{xcolor}
\usepackage{tabularx}
\usepackage{inputenc}
\usepackage{subcaption}
\newtheorem{theorem}{Theorem}
\newtheorem{definition}{Definition}
\newtheorem{lemma}{Lemma}
\newtheorem{remark}{Remark}
\newtheorem{assumption}{Assumption}
\newtheorem{example}{Example}

\title{\LARGE \bf
Approximate Stability Radius Analysis and Design in Linear Systems}

\author{Ananta Kant Rai and Vaibhav Katewa
\thanks{Ananta Kant Rai is with the Department of Electrical Communication Engineering (ECE) at the Indian Institute of Science (IISc), Bangalore. Vaibhav Katewa is with the Robert Bosch Center for Cyber-Physical Systems and the Department of ECE at IISc Bangalore. Email IDs: { \tt\small anantakant@iisc.ac.in, vkatewa@iisc.ac.in}}%
\thanks{This work is supported in part by SERB grant SRG/2021/000292.}
}

\begin{document}

\maketitle
\thispagestyle{empty}
\pagestyle{empty}

\begin{abstract}
The robustness of the stability properties of dynamical systems in the presence of unknown/adversarial perturbations to system parameters is a desirable property. In this paper, we present methods to efficiently compute and improve the approximate stability radius of linear time-invariant systems. We propose two methods to derive closed-form expressions of approximate stability radius, and use these to re-design the system matrix to increase the stability radius. Our numerical studies show that the approximations work well and are able to improve the robustness of the stability of the system.
\end{abstract}

\section{INTRODUCTION}
Stability is one of the fundamental concepts in the analysis and design of dynamical systems. Earlier, the binary notion of stability was studied and the focus was on determining whether the system is stable or not. However, in practical applications, it is crucial to assess the robustness of the system towards maintaining stability when subjected to modeling/adversarial variations in its parameters. 

To address this, the notion of Stability Radius (SR) was introduced for Linear Time-Invariant (LTI) system  $\dot{x} = Ax$ in \cite{HINRICHSEN19861}. SR quantifies the minimum-norm perturbation $\Delta$ that a stable system can tolerate before becoming unstable, and thus, it provides a quantitative measure of system stability. The perturbation forms $A+\Delta$ and $A+B\Delta C$ are referred to as unstructured and structured perturbations, respectively. In addition, $\Delta$ may also have sparsity constraints where only a subset of its entries are allowed to change. Further, the cases when $\Delta$ is allowed to be complex and real are referred to as complex SR and real SR respectively. For robust system design, SR should be taken into account while designing and deploying a system. In this paper, we present techniques to find (analysis) and improve (design) the SR of a LTI system.



\noindent \textbf{Related Work:}
Several papers have studied SR since the seminal paper \cite{HINRICHSEN19861}. Closed form expressions of the complex (unstructured and structured) SR problems were provided in \cite{HINRICHSEN19861} and \cite{HINRICHSEN1986105}. In contrast, the real SR problem is considerably difficult and no closed-form solution exists \cite{hinrichsen1990note, hinrichsen1990real}. Bounds on the real SR were obtained in \cite{qiu1991stability}, \cite{qiu1992bounds}, and a numerical computation formula was presented in \cite{qiu1993formula}. Recently, several papers have proposed numerical approaches for computing the approximate SR for Frobenius norm bounded perturbations for the non-sparse \cite{guglielmi2017approximating, zhang2023real} and sparse \cite{KATEWA2020108685, Iter_algo_2022} cases.

While the analysis problem of computing the SR has been studied extensively, not much focus has been given to the system design problem where the goal is to modify the system in order to improve its SR. The system design problem is considerably more difficult since it is a bi-level optimization problem where computation of SR appears in the constraints. Note that system design problems have been studied for other problems related to dynamical systems. For instance, the design problem is studied in the context of consensus \cite{xiao2004fast, xiao2007distributed, olfati2005ultrafast}, controllability \cite{becker2020network}, and smart grid design \cite{nagarajan2016optimal}. Further, \cite{kar2006topology, kar2008topology} present graph-based solutions to the design problems for consensus and inference. However, to the best of our knowledge, the system design problem has not yet been studied in the context of stability radius.

The main contributions of this paper are:\\
1. We present two methods to approximately compute the SR based on linear approximation of eigenvalues. The approximations enable us to get closed-form solutions which can be computed easily and efficiently.\\
2. We use the approximate SR solutions to solve the system design problem to improve the SR. The closed-form approximate solutions allow us to efficiently solve the otherwise computationally-intensive design problem.
\\
3. We present numerical simulations to show that the approximations work well, and the approximated and actual SR values are close.


  \noindent \textbf{Mathematical Notations:}$(\cdot)^{r}$ denotes the real part of a complex number. $(\cdot)^{T}$ and $(\cdot)^{*}$ denote the transpose and complex conjugate transpose of a vector/matrix, respectively. Tr$(\cdot)$ denote trace of a matrix. $\lVert \cdot \rVert$  denotes the Frobenius norm of a matrix. $\circ$ denotes the Hadamard product. $\lvert \cdot \rvert$ denotes the absoulte value. $I$ denotes the identity matrix. $j = \sqrt{-1}$ denotes the unit imaginary number. 
$\alpha(\cdot)$ denotes the spectral abscissa (maximum of the real part of all eigenvalues). $\mathbf{1}_n$ and $\mathbf{1}_{n\times m}$ denote a vector and matrix of appropriate sizes whose all elements are $1$, respectively.


\section{Problem Formulation}
\label{sec:sec_II}
We consider a perturbed continuous-time linear time-invariant (LTI) system given as
\begin{align}
\label{eq:eqn_org_dyn}
\dot{x}(t) = (A + B\Delta C)x(t),
\end{align}
where $x \in \mathbb{R}^{n}$ is the state of the system, $A \in \mathbb{R}^{n \times n}$ is the nominal system matrix, $B\in\mathbb{R}^{n \times m}$ and $C\in\mathbb{R}^{p \times n}$ are structure matrices, and $\Delta \in\mathbb{R}^{m \times p}$ is the perturbation matrix. The term $B\Delta C$ captures the perturbation to $A$, and we  denote the perturbed system as $A(\Delta) \triangleq A+B\Delta C$.

Additionally, we impose sparsity constraints on $\Delta$ as follows. Let $S \in \{0,1\}^{m \times p}$ be a binary sparsity matrix specifying the sparsity pattern as
\begin{align*}
\vspace{-100pt}
\Delta_{ij} = \begin{cases} * \quad \text{if} \quad S_{ij} = 1 , \\
0 \quad \text{if} \quad S_{ij} = 0, 
\end{cases}
\end{align*}
where $* \in \mathbb{R}$ is a scalar. The sparsity constraints are defined as 
\vspace{-10pt}
\begin{align} \label{eq:eqn_spar_const_Sr} S^{c} \circ \Delta = 0, \end{align}
where $S^{c}\triangleq 1_{m \times p } - S $ is complement of the sparsity matrix. The case where $S=1_{m \times p }$ is the non-sparse case.

\begin{remark}[Perturbation structure] The perturbation $A+B\Delta C$ can result from output feedback of linear dynamical system as: $\dot{x} = Ax +B u, y = Cx, u = \Delta y$, where $A+B\Delta C$ is the closed-loop matrix. The perturbations can also be interpreted as attacks that modify the nominal $A$ to disrupt the functioning of the system. The $B$ and $C$ matrices add structure to the perturbation (for instance, modifying certain rows or columns of $A$). The sparsity constraints allow us to change only certain entries of $A$, which is useful in networked systems where only some edge weights are allowed to change.
\end{remark}


Next, we define the stability radius corresponding to system \eqref{eq:eqn_org_dyn}, denoted by $SR(A)$. 


\begin{definition}[Stability Radius]
\begin{align}
\label{eq:eqn_SR}
     SR(A) = \min \{ \lVert \Delta \rVert: \alpha(A(\Delta)) = 0, S^{c} \circ \Delta = 0\}.
\end{align}
\end{definition}
Note that we consider $\lVert \cdot \rVert$ as the Frobenius norm throughout this paper. Without loss of generality, we assume that $A$ is stable ($\alpha(A)<0$). This ensures that $SR(A)$ is strictly positive. SR is a measure of the resiliency of the system since it captures the minimum-norm perturbation which makes the system unstable by shifting the eigenvalues of $A$ to the right-half plane. 

We propose a System Design (SD) problem where the system operator wishes to change matrix $A$ to $(A+B_o\Delta_oC_o)$ in order to improve the SR. Note that $B_o \in \mathbb{R}^{n\times m_o}, C_o \in \mathbb{R}^{p_o\times n}$ and $S_o \in \mathbb{R}^{m_o\times p_o}$ impose structure and sparsity constraints on how the system operator is allowed to modify the matrix $A$. In contrast, $B, C$ and $S$ impose constraints on ``potential" perturbations to a nominal matrix, and are used to compute the $SR$. The benign  perturbation $\Delta_{o}$ is desired to be small so that the modified system remains ``close" to the original system. The system design problem is formulated as
\begin{subequations}\label{eq:eqn_ND}
 \begin{align}
  \textbf{SD} :  \underset{\Delta_{o} \in \mathbb{R}^{m_o \times p_o}} { \text{min}}  & \lVert \Delta_{o} \rVert \tag{\ref{eq:eqn_ND}}\\
  \text{s.t.} \quad  & SR(A + B_o \Delta_{o} C_o) \geq \epsilon \label{eq:eqn_ND_cons1} \\
  & S^c_o \circ \Delta_o = 0, \label{eq:eqn_ND_cons2}
\end{align}
\end{subequations}
where $\epsilon > SR(A)$ is the increased level of $SR$ that is desired.

The \textbf{SD} problem \eqref{eq:eqn_ND} is a bi-level optimization problem since constraint \eqref{eq:eqn_ND_cons1} involves computing the SR, which itself is an optimization problem given in \eqref{eq:eqn_SR}. Note that problem \eqref{eq:eqn_SR} is non-convex and it does not admit a closed-form solution. Thus, computationally intensive iterative algorithms are used to obtain the solutions \cite{KATEWA2020108685, Iter_algo_2022, guglielmi2017approximating}. This makes the \textbf{SD} problem difficult to solve.

In order to simplify the \textbf{SD} problem, we propose to approximate the SR using approximations for the spectral abscissa of the perturbed matrix, $A(\Delta)$. Specifically, we use linear and successive-linear approximations of the spectral abscissa, denoted by $\alpha_{la}(\cdot)$ and  $\alpha_{sla}(\cdot)$, respectively. We explain these approximations later in Section \ref{sec:sec_III}. The corresponding approximated SR problems are given as:
\begin{align}
\label{eq:eqn_SR_la}
\hspace{-20pt} SR_{la}(A) = \min \{ \lVert \Delta \rVert : \alpha_{la}(A(\Delta)) = 0, S^{c} \circ \Delta = 0\}, \\
\label{eq:eqn_SR_sla}
 \hspace{-20pt} SR_{sla}(A) = \min \{ \lVert \Delta \rVert : \alpha_{sla}(A(\Delta)) = 0, S^{c} \circ \Delta = 0\}.
\end{align}

We later show that the above two problems admit closed-form solutions that can be computed easily and efficiently. 

For the system design problem, we propose to use the approximations $SR_{la}$ and $SR_{sla}$ in constraint \eqref{eq:eqn_ND_cons1}, and denote the corresponding approximate system design problems as \textbf{SD$_{la}$} and \textbf{SD$_{sla}$}, respectively. The approximate SR problems are analyzed in Section \ref{sec:sec_III} and the approximate system design problems are addressed in Section \ref{sec:sec_IV}.



\section{Approximate Stability Radius Problems}
\label{sec:sec_III}
In this section, we analyze the approximate SR problems given in \eqref{eq:eqn_SR_la} and \eqref{eq:eqn_SR_sla}. Both these problems involve linear approximation of eigenvalues, which we describe next.

\subsection{Eigenvalue Approximation}
We use the eigenvalue sensitivity analysis which dictates how eigenvalues are modified when a matrix is perturbed. In particular, we use the following result.

\begin{lemma}{(Eigenvalue Sensitivity \cite{hinrichsen2005mathematical})}
\label{lem:lem_eval_sens} 
Let $\lambda_{k}$ be a simple eigenvalue of $A$ with corresponding left and right eigenvectors, $y_{k}$ and $z_{k}$, respectively, such that $y_{k}^{*}z_{k} = 1$ for $k=1,2,\cdots, n$. Then, as $A$ is perturbed to $A(\Delta)$, the sensitivity of  $\lambda_{k}$  with respect to parameter $\Delta_{ij}$ is given as
\begin{equation}
\label{eq:eqn_sens}
\frac{\partial \lambda_{k}}{\partial  \Delta_{ij}} = y_{k}^{*} \Bigl[ \frac{\partial A(\Delta)}{\partial \Delta_{ij}}\Bigr] z_k = y_{k}^{*} B E_{ij} C z_k,
\end{equation}
where $E_{ij}$ is a matrix with $(i,j)^{\text{th}}$ entry as $1$ and all other entries as $0$.
\end{lemma}

The above lemma requires the eigenvalues to be simple, so we make the following assumption.

\begin{assumption} 
All eigenvalues of $A$ are assumed to be simple.
\end{assumption}

Let $P_k$ denote the sensitivity matrix corresponding to eigenvalue $\lambda_k$, where $[P_{k}]_{ij} = y_{k}^{*} B E_{ij} C z_k$. Based on Lemma \ref{lem:lem_eval_sens}, we approximate the real part of the eigenvalues as
\begin{align}
\label{eq:eqn_eig_approx}
\hspace{-20pt} \hat{\lambda}_{k}^r \approx \lambda_k^r + \sum_{i=1}^m \sum_{j=1}^p [P_k^r]_{ij} \Delta_{ij} = \lambda_k^r + \mathbf{1}_m^T(P_k^r\circ \Delta) \mathbf{1}_p, 
\end{align}
where $\lambda_k^r$ and $P_k^r$ denote the real parts of $\lambda_k$ and $P_k$ respectively.

\begin{remark}[Effect of normality of $A$ on the approximations] \label{rem:normal_mat}
    When $A$ is normal, the sensitivity values in \eqref{eq:eqn_sens} are small \cite{hinrichsen2005mathematical}, and as a result, the approximation provided in \eqref{eq:eqn_eig_approx} works well. In contrast, for non-normal matrices, the approximation error may be large. We comment on this fact later in the simulation Section \ref{sec:Section_Num_Sim}.
\end{remark}

\subsection{Stability Radius via Linear Approximation}


Based on the linear approximation (LA) of the eigenvalues in \eqref{eq:eqn_eig_approx}, we re-write \eqref{eq:eqn_SR_la} as
\begin{multline}
SR_{la}(A) = \min \{ \lVert \Delta \rVert : \\ \underset{k = 1,...,n}{\max}\{\lambda_{k}^{r} + \mathbf{1}_{m}^{T}(P_{k}^{r}\circ \Delta) \mathbf{1}_{p} \} = 0, S^{c} \circ \Delta = 0\} \label{eq:eqn_SR_approx},
\end{multline}
\vspace{-15pt}
\begin{multline}
    = \underset{k = 1,...,n}{\min} \{\lVert \Delta_{k}^{*} \rVert\}, \quad \text{where} \\
 \Delta_{k}^{*}  = \arg \min \{\lVert \Delta \rVert : \\\lambda_{k}^{r} + \mathbf{1}_{m}^{T}(P_{k}^{r}\circ \Delta) \mathbf{1}_{p} = 0, S^{c} \circ \Delta = 0\}.\label{eq:eqn_delt_k_opt_la}
\end{multline}
Thus, we solve the optimization problem \eqref{eq:eqn_delt_k_opt_la} individually for each eigenvalue, and then take the minimum-norm over these solutions to get $SR_{la}(A)$. Note that problem \eqref{eq:eqn_delt_k_opt_la} is a quadratic optimization problem with linear equality constraint and admits a closed-form solution that can be computed quickly.

Next, we discuss the feasibility of the $SR_{la}$ problem.

\begin{lemma}{(Feasibility)}\label{lem:lem_feasibility_la}
     The optimization problem in \eqref{eq:eqn_SR_approx} is feasible if and only if  $ (S \circ P_{k}^{r}) \neq 0$ holds true for at least one $k\in\{1,2,\cdots,n\}$.
\end{lemma}
\proof
In \eqref{eq:eqn_delt_k_opt_la}, the second equality constraint $S^{c} \circ \Delta = 0$ is equivalent to  $\Delta = S \circ \bar{\Delta}$, where $\bar{\Delta}$ is any arbitrary matrix. Substituting $\Delta$ in the first equality constraint in \eqref{eq:eqn_delt_k_opt_la}, we get 
\begin{align*}
     \mathbf{1}_{m}^{T}(P_{k}^{r}\circ S \circ \bar{\Delta}) \mathbf{1}_{p}  = - \lambda_{k}^{r}.
\end{align*} 
Since $\alpha(A)<0$, we have $\lambda_{k}^{r}\neq 0$. Thus, a solution $\bar{\Delta}$ exist for the above equation if and only if $P_{k}^{r}\circ S \neq 0$. The result then follows from \eqref{eq:eqn_SR_approx}. \hfill $\blacksquare$

Note that the condition $(P_k^{r} \circ S) = 0$ implies that under $S^{c} \circ \Delta=0$, the term $\mathbf{1}_{m}^{T}(P_{k}^{r}\circ \Delta) \mathbf{1}_p=0$. Therefore, the $k^{\text{th}}$ eigenvalue of $A$ cannot be shifted by the perturbation with the given sparsity constraints. If this holds true for all eigenvalues, then none of the eigenvalues of $A$ can be shifted, and $SR_{la}(A) = \infty$.
\begin{example} Consider $A = \begin{bmatrix}
    -1 & 0.5 \\ -2 & 0.2
\end{bmatrix}, B = \begin{bmatrix}
      0 & 1 \\ 0 & 1
\end{bmatrix}, \\ C = \begin{bmatrix}
    0.4  & 1\\ 1 & 1
\end{bmatrix}$ and $
S = \begin{bmatrix}
    1 & 1 \\ 0 & 0
\end{bmatrix}$. Here, $\lambda = -0.4 \pm 0.8j$, $P_1^{r} = P_2^{r} = \begin{bmatrix}
    0 &  0 \\ 0.7 & 1
\end{bmatrix}.$ Thus, $P_1^r \circ S = P_2^{r} \circ S = 0$. Hence, the feasibility condition is violated and none of the eigenvalues can be shifted.
\end{example}

Next, instead of solving problem \eqref{eq:eqn_SR_approx} directly, we present an equivalent alternate reformulation of \eqref{eq:eqn_SR_approx}. This reformulation will also be required in the next subsection for the successive- linear approximation of the spectral abscissa, and is given by
 \begin{align}
\lVert \Delta_k^{*} \rVert = \{ \beta: SA_k(\beta) = 0\} \label{eq:eqn_del_at_sak_0}, \text{ where}
\end{align}
\vspace{-15pt}
\begin{subequations}\label{eq:eqn_sr_la_prob}
\begin{align}
SA_k(\beta) = \underset{\Delta}{\max} \quad & \lambda_{k}^{r} + \mathbf{1}_{m}^{T}(P_{k}^{r}\circ \Delta) \mathbf{1}_{p} \tag{\ref{eq:eqn_sr_la_prob}}\\ 
\text{s.t} \quad & \lVert \Delta \rVert \leq \beta \label{eq:eqn_del_norm_beta}\\ \label{eq:eqn_sparsity_const}
& S^{c} \circ \Delta = 0.
\end{align} 
\end{subequations}

Note that in both problems \eqref{eq:eqn_SR_approx} and \eqref{eq:eqn_del_at_sak_0}, we compute the approximate spectral abscissa of the perturbed system and determine the minimum-norm perturbation that shifts it to the unstable region. Hence, the solutions to these two problems are identical.

Problem \eqref{eq:eqn_sr_la_prob} is convex with linear cost, and quadratic and linear constraints. Next, we present its closed-form solution.

\begin{theorem}
Let the feasibility condition in Lemma \ref{lem:lem_feasibility_la} holds true, and define $\mathcal{K} = \{k: (S \circ P_k^{r}) \neq 0\}$. Then, the solution of the optimization problem \eqref{eq:eqn_del_at_sak_0} is given by 
\begin{align}
    \Delta_k^{*} = -\frac{\lambda_{k}^{r} (S \circ P_k^{r})}{ \lVert S \circ P_k^{r} \rVert^{2}}, \label{eq:eqn_Delta_opt} \quad  k \in \mathcal{K}. 
\end{align}
Further, we have
\begin{align}
       SR_{la}(A) = \underset{k \in \mathcal{K}}{\min} \biggl \{\frac{-\lambda_k^{r}} {\lVert S \circ P_k^{r} \rVert} \biggr \}. \label{eq:eqn_SR_la_soln}
   \end{align}
\end{theorem}
\proof
We first solve optimization problem \eqref{eq:eqn_sr_la_prob} for $k \in \mathcal{K}$  using the first-order KKT conditions. Using $\lVert A \rVert^{2} = \text{Tr}(A^{T}A),$ \eqref{eq:eqn_del_norm_beta} can be rewritten as
\begin{align*}
 \lVert \Delta \rVert \leq \beta \Leftrightarrow \text{Tr}(\Delta^{T}\Delta) \leq \beta^{2}.   
\end{align*}

Let $l \leq 0$ and $M \in \mathbb{R}^{m \times p}$ be the Lagrangian multipliers associated with constraints \eqref{eq:eqn_del_norm_beta} and \eqref{eq:eqn_sparsity_const}, respectively. The Lagrangian function is given by 
\begin{align*}
    \hspace{-20pt} \mathcal{L} =  \lambda_k^{r} +  \mathbf{1}_{m}^{T}(P_{k}^{r}\circ \Delta) \mathbf{1}_{p} + l(\lVert \Delta \rVert - \beta) \\ + \mathbf{1}_{m}^{T}(M \circ (S^{c} \circ \Delta)) \mathbf{1}_{p}, \\
    = \lambda_k^{r} +  \text{Tr}((P_{k}^{r})^{T} \Delta) + l\:(\text{Tr}(\Delta^{T}\Delta) - \beta^{2}) \\ + \text{Tr}((M \circ S^{c})^{T}\Delta),
\end{align*}
where we have used the property $\mathbf{1}_{m}^{T}(A\circ B) \mathbf{1}_{p} = \text{Tr}(A^T B)$. Differentiating $\mathcal{L}$ with respect to $\Delta$ and equating to 0, we get
\vspace{-10pt}
\begin{align}
    \frac{\partial \mathcal{L}}{\partial \Delta} =  P_{k}^{r} + 2l\Delta + 
    M\circ S^{c}= 0. \label{eq:eqn_dl_ddel}
\end{align}
Taking the Hadamard product of \eqref{eq:eqn_dl_ddel} with $S^{c}$, and using \eqref{eq:eqn_spar_const_Sr}  and $ S^{c} \circ S^{c} = S^{c}$, we get
\begin{align*}
&S^{c} \circ P_{k}^{r} + 2l(S^{c} \circ \Delta) + S^{c} \circ ( M\circ S^{c}) = 0,\\
     \Rightarrow  &M\circ S^{c}  = - S^{c} \circ P_{k}^{r}.  
\end{align*}
Substituting the above expression in \eqref{eq:eqn_dl_ddel} and using $S = 1_{m\times p} - S^{c}$, we get
\begin{align}
\mathbf{1}_{m \times p} \circ P_{k}^{r} + 2l\Delta - S^{c} \circ P_k^{r}= 0, \nonumber \\
 \Rightarrow 2l\Delta = -S \circ P_k^{r}. \label{eq:eqn_delta_del_step}
\end{align}
Since $(S \circ P_k^r) \neq 0$ for $k \in \mathcal{K}$, \eqref{eq:eqn_delta_del_step} implies that $l \neq 0$, and we get
\begin{align}
    \Delta = -(S \circ P_k^{r})/2l. \label{eq:eqn_delt_opt_k}
\end{align}
By complementary slackness, $l \neq 0$ implies that constraint \eqref{eq:eqn_del_norm_beta} is active. 
Substituting $\Delta$ obtained in \eqref{eq:eqn_delt_opt_k} in the active constraint $\lVert \Delta \rVert = \beta$, we get
\begin{align*}
   \hspace{-10pt} l = - \lVert S \circ P_k^{r} \rVert /2\beta. \quad \text{(Since $l\leq 0$ from dual feasibility)}
\end{align*}
Next, substituting $l$ in \eqref{eq:eqn_delt_opt_k}, we get the solution to Problem \eqref{eq:eqn_sr_la_prob} as:
\begin{align}
     \Delta_{k} = \frac{\beta (S \circ P_k^{r})}{\lVert S \circ P_k^{r} \rVert} \label{eq:eqn_del_k_opt},
\end{align}
 and $SA_k(\beta) = \lambda_k^{r} + \beta \lVert S \circ P_k^{r} \rVert $.  Since Problem \eqref{eq:eqn_sr_la_prob} is convex, any solution satisfying the first-order KKT conditions is a global minimum. 
 Next, we solve the equality in \eqref{eq:eqn_del_at_sak_0} to get
 \vspace{-10pt}
\begin{align}
\label{eq:eqn_beta_opt}
    \lVert \Delta_k^{*} \rVert = \beta^{*} = - \frac{\lambda_{k}^{r}}{\lVert S \circ P_k^{r}\rVert},
 \end{align}
 and substitute $\beta^{*}$ in \eqref{eq:eqn_del_k_opt} to get \eqref{eq:eqn_Delta_opt}
 \begin{align*}
    \Delta_k^{*} = -\frac{\lambda_{k}^{r} (S \circ P_k^{r})}{\lVert S \circ P_k^{r} \rVert^{2}}.
 \end{align*}
 The result \eqref{eq:eqn_SR_la_soln} then follows from \eqref{eq:eqn_delt_k_opt_la}. 
\hfill $\blacksquare$

Result \ref{eq:eqn_beta_opt} implies that $SR_{la}$ is dependent on the ratio $\lambda_{k}^{r}/\lVert (S \circ P_k^{r})\rVert$. Intuitively, if this ratio is large $SR_{la}$ is large. This ratio becomes large when $\lambda_k^r$ is large and $\lVert S \circ P_k^{r} \rVert$ is small and vice versa. Thus, system with large eigenvalues and corresponding small sensitivities require large perturbation to shift.

Our approach of linear approximation of eigenvalues  works well if the perturbations are small. Intuitively, if $SR_{la}(A)$ is small, then we expect the approximation to work well, and $SR_{la}(A)$ to be close to the actual $SR(A)$. However, in the case when $SR_{la}(A)$ is large, the linear approximation may not be precise. To address this issue, we propose a successive-linear approximation (SLA) approach next.



\subsection{Stability Radius via Successive-Linear Approximation}

As mentioned before, we expect the linear approximation to work well if the perturbations are small. Motivated by this, we propose to decompose the perturbation $\Delta$ as
\begin{align}
\label{eq:eqn_delta_Sla}
\Delta = \Delta^{(1)} + \Delta^{(2)} + \cdots + \Delta^{(J)},
\end{align}
and ensure that each perturbation is small, that is, $\lVert \Delta^{(j)} \rVert \leq \beta << 1$. We compute the optimal value of $\Delta^{(j)}$ (denoted by $\Delta^{(j,*)}$) in a successive/iterative manner and this results in a successive approximation of  the spectral abscissa.

Let $A_{j-1} \triangleq A+ B\Delta^{(1,*)}C + B\Delta^{(2,*)}C + \cdots + B\Delta^{(j-1,*)}C$ with $A_0 \triangleq A$.
Next, we explain the steps to obtain $\Delta^{(j,*)}$.

1. We solve problem \eqref{eq:eqn_sr_la_prob} with $\lambda_k$ and $P_k$ being the eigenvalues and sensitivity matrix of $A_{j-1}$. Let the optimal solution to this problem be denoted by $\Delta_{k}^{(j,*)}$.

2. Compute
\begin{align}
\Delta^{(j,*)} = \underset{k = 1,\cdots,n}{\arg \max} \{ \alpha(A_{j-1}+B\Delta_{k}^{(j,*)}C) \}.\label{eq:eqn_delt_j_opt}
\end{align}

3. Update $A_j = A_{j-1}+B\Delta^{(j,*)}C$.

4. Repeat the above steps until $\alpha(A_j)<0$.

Let $J$ denote the number of iterations of the above algorithm. Then,
\begin{align}
SR_{sla}(A) = \lVert \Delta^{(1,*)} + \Delta^{(2,*)} + \cdots + \Delta^{(J,*)} \rVert. \label{eq:eqn_sr_sla_sum}
\end{align}

The algorithm is as follows:
\begin{algorithm}
\caption{SR via successive-linear approximations}\label{alg:Sr_Algo_sla}
\begin{algorithmic}
\Require $A, B, C, S, \beta$
\State \textbf{Output:} $SR_{sla}(A)$
\While{$\alpha(A_{j}) < 0$} \Comment{$\alpha(A_{j}) > 0 \Rightarrow $ unstable region} 
\For{$k = 1,...,n$}
\State $P_{k} \gets \eqref{eq:eqn_sens}$ \Comment{sensitivity matrix of $\lambda_{k}$}
\State $\Delta_{k}^{(j,*)}\gets \eqref{eq:eqn_del_k_opt}$
\EndFor
\State $\Delta^{(j,*)} \gets \eqref{eq:eqn_delt_j_opt}$ 
\State $A_j = A_{j-1}+B\Delta^{(j,*)}C$
\EndWhile
\end{algorithmic}
\end{algorithm} 

Several remarks are in order. First, note that although computation of $SR_{sla}(A)$ is an iterative procedure, in each iteration we use the closed-form expression given in \eqref{eq:eqn_del_k_opt}. Thus, the overall computation time is small. Second, the algorithm is greedy in nature since at each iteration in \eqref{eq:eqn_delt_j_opt} we are picking a perturbation that corresponds to the largest spectral abscissa. Third, since we are using the eigenvalue sensitivity result given in Lemma \ref{lem:lem_eval_sens}, our approach requires that in each iteration, all eigenvalues of $A_j$ are simple. Fourth, we conjecture that the $SR_{sla}(A)$ is a better approximation than $SR_{la}(A)$ since the former involves several small perturbations as compared to a single but potentially large perturbation in the latter. Our conjecture is supported by simulations presented in simulation Section \ref{sec:Section_Num_Sim}.
\begin{remark} [Implementation details of Algorithm \ref{alg:Sr_Algo_sla}]
Each iteration of Algorithm \ref{alg:Sr_Algo_sla} involves solving Problem \ref{eq:eqn_sr_la_prob} for $A_j$. This requires Problem \ref{eq:eqn_sr_la_prob} to be feasible as per the condition provided in Lemma \ref{lem:lem_feasibility_la}. If this feasibility condition is violated, we slightly perturb $A_j$ randomly such that the problem becomes feasible and continue thereafter.

Also, for Algorithm \ref{alg:Sr_Algo_sla} to terminate, we require that $\alpha(A_j) > \alpha(A_{j-1})$ holds true at each iteration. However, since $A_j$ is computed based on linear approximation of eigenvalues, this condition might be violated. In this case, then we repeat the iteration with a slightly higher value of $\beta$ such that $\alpha(A_j) > \alpha(A_{j-1})$ holds, and continue thereafter. However, we remark that in our simulation studies, we did not encounter these issues.
\end{remark}

\begin{remark} [Comparison of LA and SLA based approaches] \label{rem:Comp_LA_SLA}
The SLA-based approach to compute $SR_{sla}$ is more accurate than the LA-based approach to compute $SR_{la}$. However, the former is iterative in nature, and therefore, requires more computational time as compared to the latter (details are presented in Section \ref{sec:Section_Num_Sim}). Thus, depending on the accuracy requirements and computational resources, one can select one of these two approaches.
\end{remark}

\begin{remark}[Approximate solutions can aid other SR algorithms]
    The approximate solutions provided by our algorithms can serve as a good initializing point for the iterative algorithms proposed earlier \cite{hinrichsen1990real, KATEWA2020108685, Iter_algo_2022} for the SR problem. This can considerably reduce the execution time of these algorithms. We defer the demonstration of this as a future work.
\end{remark}

\section{System Design Problems}
\label{sec:sec_IV}
In this section, we study the system design problem mentioned in $\eqref{eq:eqn_ND}$.  We wish to find a benign perturbation $\Delta_{o}$ added by the operator that improves the stability radius of the system. In problem \eqref{eq:eqn_ND}, the constraint \eqref{eq:eqn_ND_cons1} involves computation of $SR(A + B_o\Delta_o C_o)$. As mentioned earlier, computation of the actual stability radius is computationally difficult and requires an iterative procedure. Hence, we use the approximated SR - $SR_{la}(A)$ and $SR_{sla}(A)$ presented in Section \ref{sec:sec_III} to approximate \eqref{eq:eqn_ND_cons1} in the system design problem. 

\subsection{System Design via Linear Approximation}
We replace $SR(\cdot)$ by $SR_{la}(\cdot)$ in constraint \eqref{eq:eqn_ND_cons1} to get
\begin{subequations}\label{eq:eqn_ND_la}
 \begin{align}
   \underset{\Delta_{o} \in \mathbb{R}^{m_o \times p_o}} {\text{min}} & \lVert \Delta_{o} \rVert \tag{\ref{eq:eqn_ND_la}}\\
  \text{s.t.} \quad  & SR_{la}(A + B_o \Delta_{o} C_o) \geq \epsilon \label{eq:eqn_ND_la_cons1} \\
 & S^c_o \circ \Delta_o = 0. \label{eq:eqn_ND_la_cons2}
\end{align}
\end{subequations}

Next, we focus on the term $SR_{la}(A + B_o \Delta_{o} C_o)$. Note that in the $SR_{la}$ problem, the optimal solution $\Delta_k^{*}$ in \eqref{eq:eqn_delt_k_opt_la} corresponds to matrix $A$, where $\lambda_k$ and $P_k$ are eigenvalue and sensitivity matrix of $A$. Similarly, let us denote the corresponding optimal solution for the matrix $A + B_o \Delta_{o} C_o$ as $\Delta_{k,o}^{*}$. Then, by \eqref{eq:eqn_delt_k_opt_la}, we have
\begin{align}
 SR_{la}(A + B_o \Delta_{o} C_o) = \underset{k = 1,\cdots, n}{\min} \{ \lVert \Delta_{k,o}^{*} \rVert \}.\label{eq:eqn:sr_la_delta_o}
\end{align}
Using \eqref{eq:eqn:sr_la_delta_o}, we reformulate Problem \eqref{eq:eqn_ND_la} as:
\begin{align}
     \textbf{SD}_{la}: \underset{\Delta_{o} \in \mathbb{R}^{m_o \times p_o}}{\text{min}} & \lVert \Delta_{o} \rVert \label{eq:eqn_ND_la_final}\\
 \hspace{20 pt}  \text{s.t.} \quad & \lVert \Delta_{1,o}^{*} \rVert \geq \epsilon \nonumber\\ 
  & \lVert \Delta_{2,o}^{*} \rVert \geq \epsilon \nonumber \\
  \hspace{100pt} \vdots \nonumber \\
  & \lVert \Delta_{n,o}^{*} \rVert \geq \epsilon \nonumber \\
 & S_o^{c} \circ \Delta_o = 0 \nonumber.
\end{align}

Thus, the single constraint in \eqref{eq:eqn:sr_la_delta_o} which involves a $\min(\cdot)$ function is converted into multiple constraints. This is desirable since the $\min(\cdot)$ function can be non-smooth and can cause numerical difficulties during the optimization problem.

Several comments are in order. First, \textbf{SD}$_{la}$ is not a bi-level optimization problem since $\Delta_{k,o}^{*}$ in the inequality constraints admit a closed-form solution. Thus, it is computationally tractable. Second, the terms $\Delta_{k,o}^{*}$ in the inequality constraints depend on $\Delta_o$ since their computation depends on the eigenvalues and sensitivity matrices of $A+B_o\Delta_o C_o$. This dependence makes \textbf{SD}$_{la}$ a non-convex problem and it may have multiple local minima. Third, we solve the \textbf{SD}$_{la}$ problem using numerical solvers. More details are presented in Section \ref{sec:Section_Num_Sim}.

\subsection{System Design via Successive-Linear Approximation}
As mentioned earlier, $SR_{sla}$ gives us a better approximation of stability radius as compared to $SR_{la}$. Therefore, we now approximate the system design problem by replacing $SR(\cdot)$ by $SR_{sla}(\cdot)$ in \eqref{eq:eqn_ND} to get
\begin{subequations}\label{eq:eqn_ND_sla}
 \begin{align}
   \textbf{SD}_{sla:} \underset{\Delta_{o} \in \mathbb{R}^{m_o \times p_o}} {\text{min}}  & \lVert \Delta_{o}\rVert \tag{\ref{eq:eqn_ND_sla}}\\
  \text{s.t.} \quad & SR_{sla}(A + B_o \Delta_{o} C_o) \geq \epsilon \label{eq:eqn_ND_sla_cons1} \\
 & S^c_o \circ \Delta_o = 0.
\end{align}
\end{subequations}

We use \eqref{eq:eqn_sr_sla_sum} to compute $SR_{sla}(\cdot)$ in constraint \eqref{eq:eqn_ND_sla_cons1}, and note that this computation uses iterative closed-form expressions. Further, similar to \textbf{SD}$_{la}$ problem, \textbf{SD}$_{sla}$ problem is also non-convex and is solved using numerical solvers. More details are presented in Section \ref{sec:Section_Num_Sim}.

\section{Numerical Simulations}\label{sec:Section_Num_Sim}
In this section, we present numerical simulation results of our algorithms. We perform the simulations using MATLAB R2023a. We first consider the spectral abscissa and SR problems, and later the system design (\textbf{SD}) problems. 

\subsection{Approximations of Spectral Abscissa and SR}
Our SR algorithms rely crucially on approximations of the spectral abscissa. Thus, we first analyse the quality of these approximations. Recall that $\alpha(\cdot), \alpha_{la}(\cdot)$ and $\alpha_{sla}(\cdot)$ denote the spectral abscissas without approximation, with linear approximation and with successive-linear approximations, respectively. For a given norm bound $\gamma$, we compute the spectral abscissas by solving the following problems 
\begin{align}
 \hspace{-4pt} \alpha (\gamma) = \underset{\Delta}{\max} \{\alpha(A(\Delta)) : \lVert \Delta \rVert \leq \gamma, S^{c} \circ \Delta = 0\}, \label{eq:eqn_abs} \\
  \hspace{-12pt} \alpha_{la} (\gamma) = \underset{\Delta}{\max} \{\alpha_{la}(A(\Delta)) : \lVert \Delta \rVert \leq \gamma, S^{c} \circ \Delta = 0\}, \label{eq:eqn_abs_la} \\
 \hspace{-18pt}   \alpha_{sla} (\gamma) = \underset{\Delta}{\max} \{\alpha_{sla}(A(\Delta)) : \lVert \Delta \rVert \leq \gamma, S^{c} \circ \Delta = 0\} \label{eq:eqn_abs_sla}.
  \end{align}
  
We solve Problem \eqref{eq:eqn_abs} by performing an exhaustive grid search over the set of perturbations that satisfy the constraints in \eqref{eq:eqn_abs}. Problem \eqref{eq:eqn_abs_la} it equivalent to Problem \eqref{eq:eqn_sr_la_prob} for the $k^{\text{th}}$ eigenvalue. Thus, we have $\alpha_{la}(\gamma) = \underset{k = 1, \cdots n}{\max} \{SA_{k}(\gamma)\}$. The computation of $\alpha_{sla}(\gamma)$ in \eqref{eq:eqn_abs_sla} is done in a successive manner similar to Algorithm \ref{alg:Sr_Algo_sla}. The difference is that we run the algorithm until $\lVert \Delta^{(1,*)} + \Delta^{(2,*)} + \cdots + \Delta^{(j,*)} \rVert \leq \gamma$ is not violated (see \eqref{eq:eqn_delta_Sla}). Let $J_{\gamma}$ denote the iteration number until the above condition in not violated. Then, we get $\alpha_{sla}(\gamma) = \alpha(A + B(\Delta^{(1,*)} + \Delta^{(2,*)} + \cdots + \Delta^{(J_{\gamma},*)})C).$
  
  
We define the approximation errors of the spectral abscissas as $e_{la}(\gamma) = \lvert \alpha(\gamma) - \alpha_{la}(\gamma) \rvert$ and $e_{sla} (\gamma) = \lvert \alpha(\gamma) - \alpha_{sla}(\gamma) \rvert$. Further, we use the following non-normality measure (normality gap \cite{elsner1987measures}) of matrix $A: NG(A) = \lVert A^{T}A - AA^{T} \rVert$. To begin, we consider two cases, one with a normal $A$ and the other with a non-normal $A$.

\textbf{Case I} (Normal $A$, $NG(A) = 0$):
\begin{align*}
     A &= \begin{bmatrix} -1.2 & -0.3 & -1 \\ -0.3 & -1.4 & -1 \\ -1 & -1  & -1.3 
\end{bmatrix}, & B &= \begin{bmatrix}
    0.4 & 0.1 \\ 0.2 & 0.3 \\ 0.4 & 0.1
\end{bmatrix}, \\
    C &= \begin{bmatrix}
    0.7 & 0.3 & 0.3 \\ 0.1 & 0.3 & 0.6
\end{bmatrix}, &   S &= \begin{bmatrix}
    1 & 0 \\ 0 & 1
\end{bmatrix}.
\end{align*}
\textbf{Case II} (Non-Normal $A$, $NG(A) = 148.29$):
\begin{align*}
     A & = \begin{bmatrix}
    -3& -4 & -7 \\
    -1 & -9 & -6 \\
    -1 & -1  & -9
\end{bmatrix}, & B &= \begin{bmatrix}
    1.3 & 1 \\
    1 & 0.7 \\
    0.5 & 1.4 \\
\end{bmatrix}, \\
    C &= \begin{bmatrix}
    1& 0.8 & 1.3 \\
    1.5 & 1.8 & 0.8
\end{bmatrix}, &   S &= \begin{bmatrix}
    1 & 0 \\ 0 & 1
\end{bmatrix}.
\end{align*}
\begin{figure}[thpb]
\begin{subfigure}{\columnwidth}
    \centering
    \includegraphics[width=\columnwidth]{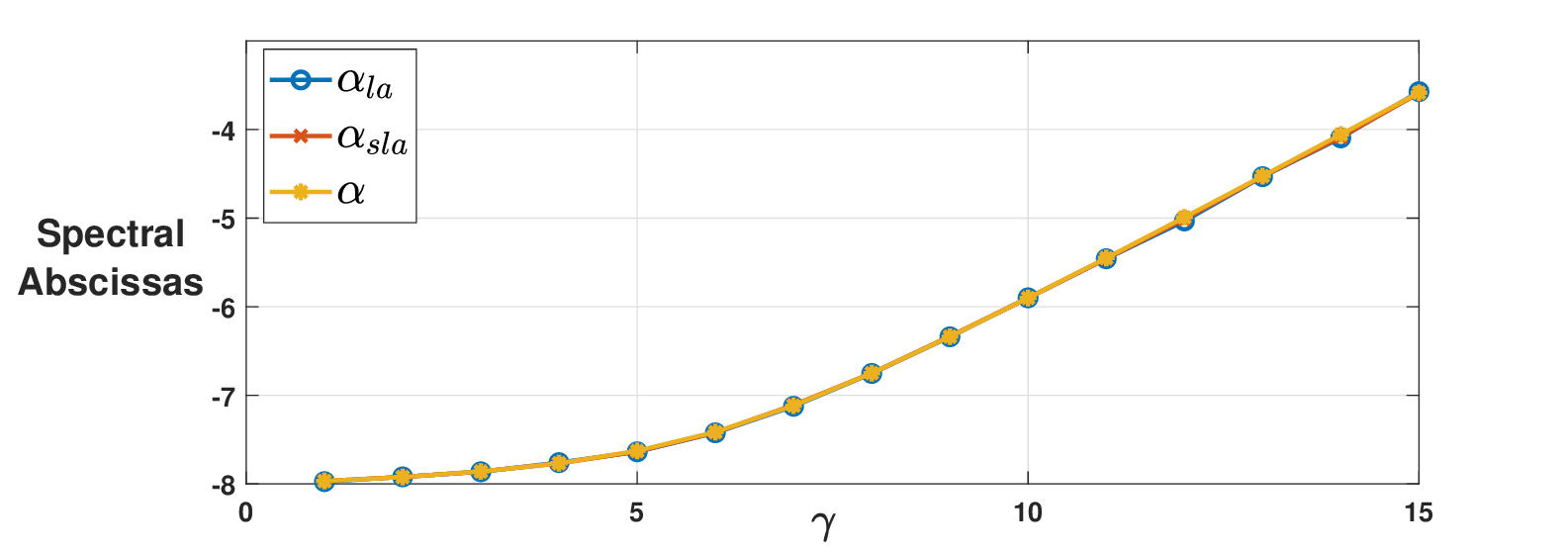}
    \caption{Case I: A is normal}
    \label{fg:fig_sr_normal}
\end{subfigure}
\begin{subfigure}{\columnwidth}
    \centering
    \includegraphics[width=\columnwidth]{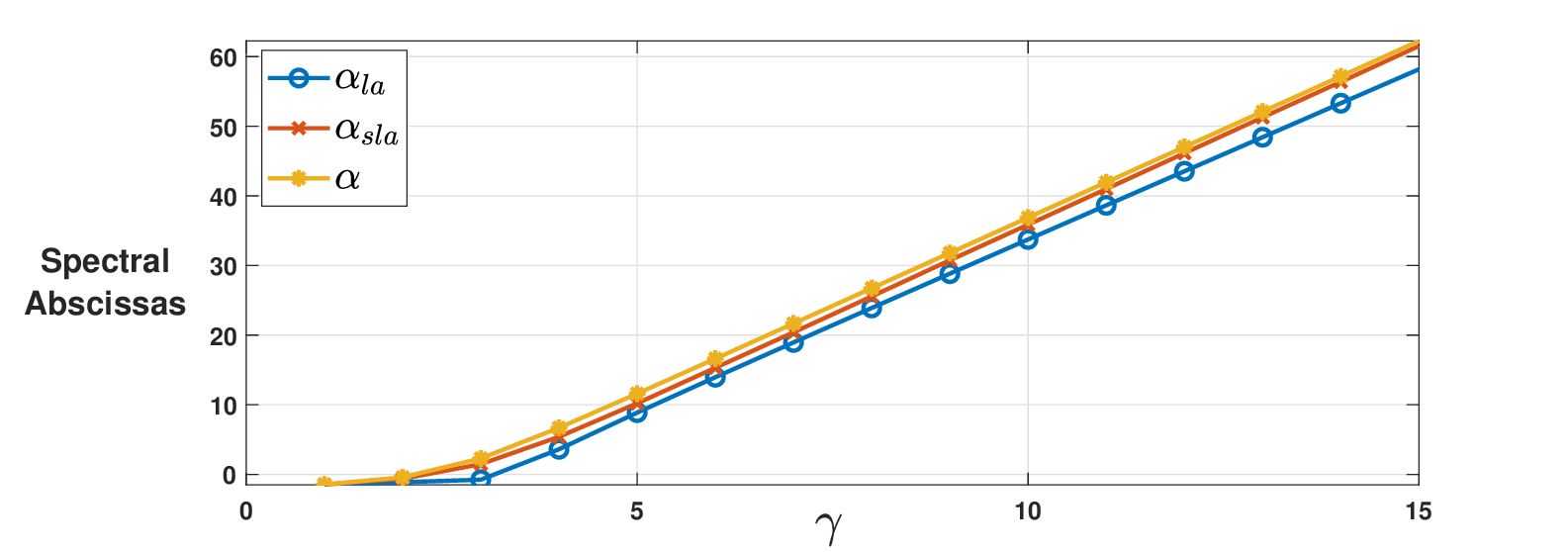}
    \caption{Case II: A is non-normal}
    \label{fg:fig_sr_non_normal}
\end{subfigure}
\caption{Variation of spectral abscissas as a function of $\gamma$.}
\label{fg:fig_abs_sr}
\end{figure}

Figure \ref{fg:fig_abs_sr} shows the plots of $\alpha(\gamma), \alpha_{la}(\gamma)$ and $\alpha_{la}(\gamma)$ for the above two cases. For Case I in Figure \ref{fg:fig_sr_normal}, we observe that $\alpha_{la}(\gamma)$ and $\alpha_{sla}(\gamma)$ overlap with $\alpha(\gamma)$ for all values of $\gamma$. This is because $A$ is normal, and hence, the eigenvalue sensitivities are small ($\lVert P_1^r \rVert = 0.6063,  \lVert P_2^r \rVert = 0.0666, \lVert P_3^r \rVert = 0.0399$). Thus, the approximations work well. On the other hand, for Case II in Figure \ref{fg:fig_sr_non_normal}, we observe differences between $\alpha_{la}(\gamma), \alpha_{sla}(\gamma)$ and $\alpha(\gamma)$, especially for large values of $\beta$. This is because $A$ is non-normal, and hence, the eigenvalue sensitivities are large ($\lVert P_1^{r} \rVert = 8.3881, \lVert P_2 ^{r} \rVert = 0.7848,  \lVert P_3^{r} \rVert = 1.9765$). However, we observe that $\alpha_{sla}(\gamma)$ provides a better approximation than $\alpha_{la}(\gamma)$. Further, as $\gamma$ increases, spectral abscissas increase, this is due to the fact that as perturbation norm increases, eigenvalue spectrum increases, hence, the spectral abscissas increase.



 \begin{figure}[thpb]
      \centering
\includegraphics[width=\columnwidth]{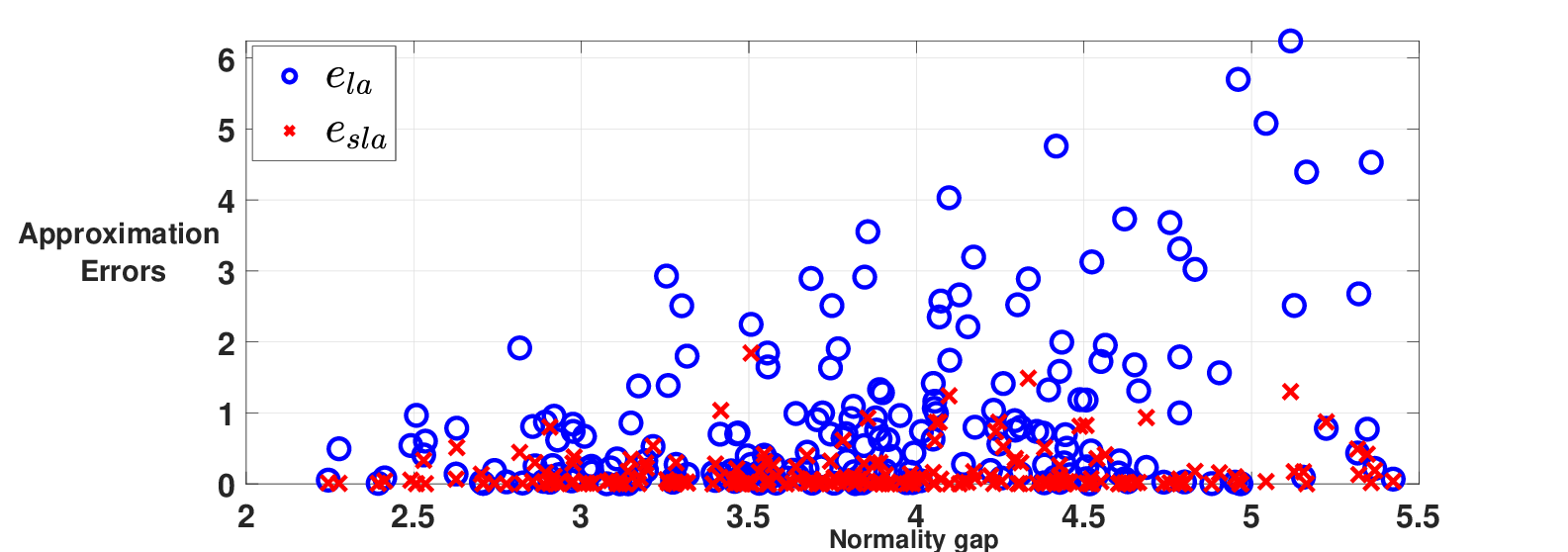}
      \caption{Variation of approximation errors $e_{la}$ and $e_{sla}$ as a function of normality gap $NG$ for 200 random triplets $(A,B,C)$. }
      \label{fg:fig_sr_random}
   \end{figure}
Next, we perform a similar comparison for a set of 200 random triplet of matrices $(A,B,C)$ with $n=5, m=2$ and $p=2$, $S = \begin{bmatrix}
    1 & 0 \\ 0 & 1
\end{bmatrix}$ and $\gamma = 10$.
For each triplet, we compute normality gap $NG(A)$ and approximation errors $e_{la}(\gamma)$ and $e_{sla}(\gamma)$. Figure \ref{fg:fig_sr_random} shows that the errors $e_{sla}(\gamma)$ are considerably smaller than $e_{la}(\gamma)$, implying that the successive-linear approximation performs better than the linear approximation. Further, $e_{la}(\gamma)$ is larger for higher values of the normality gap, whereas $e_{sla}(\gamma)$ remains small. This shows that the SLA-based algorithm is more suited for non-normal problems.

These results provide an empirical evidence that our approximation based approaches are well suited for approximate stability radius and system design problems.

Next, we use our algorithms to compute the approximate SR values $SR_{la}$ and $SR_{sla}$ for a subset of test problems in the COMPl$_{e}$ib \cite{leibfritz2006compleib}. This library contains test problems for LTI  control systems and $A, B,C$ matrices are specified for each problem. However, sparsity pattern is not specified for any of the test problems in the library. For some problems, we define the sparsity pattern and mark ($S$) against those in Table \ref{Tb:Tbl_SR_vals}.
\begin{table}[h]
\begin{center}
\caption{SR values for different test problems}
\scalebox{0.95}{\begin{tabular}{|c|c|c|c|c|c|c|}
\hline
 Test Problem & $n$ & $SR$ & $SR_{la}$ & $SR_{sla}$ & $\tau_{la}$ & $\tau_{sla}$ \\
 \hline
  Case I \cite{KATEWA2020108685} & 4 & 0.5159 &  0.5218 & 0.5140 & 0.0030& 0.0959\\
 \hline
 Case II \cite{KATEWA2020108685} & 4 & 0.5653 & 0.6110 & 0.5694  &0.0035 & 0.0929\\
 \hline
 HF2D12 &5 &1.4912  & 1.5371  &1.3921 &  0.0017 & 0.1591 \\
 \hline
 HF2D13 (S) & 5 & 0.0424 & 0.0421 & 0.0422 & 0.0057 & 0.0274\\
 \hline
 TG1 & 10 & 0.0673  & 0.0642 & 0.0661 & 0.0014 & 0.0022\\
 \hline
 AGS & 12 & 0.0688 & 0.0624& 0.0719  & 0.0022 & 0.534\\
 \hline
 WEC2 (S) & 10 & 0.0435& 0.0420 & 0.0430  & 0.0035 & 0.0454\\
 \hline
 WEC3 & 10  &  0.5534 & 0.5221 & 0.5410 & 0.0052 & 0.0237\\
 \hline
 BDT1 & 11 & 0.0515 & 0.051 & 0.0514 & 0.0030 & 0.0099\\
 \hline
 MFP & 4& 0.7986 & 0.8123 & 0.8011 & 0.0020 & 0.1211\\
 \hline
 UWV & 8 & 0.127 & 0.132 & 0.1239 & 0.0031 & 0.0123\\
 \hline
 EB1 & 10 & 0.0201 & 0.0205 & 0.0200 &0.0137 & 0.1249\\
 \hline
 PSM (S) & 7& 0.4432 & 0.3508 & 0.4190 & 0.0198 & 0.4074 \\
 \hline
 CDP (S) & 120 & 0.0073 & 0.0071 & 0.0074 & 0.0649 & 0.3550\\
 \hline
\end{tabular}}\label{Tb:Tbl_SR_vals} 
\end{center}
\footnotesize{Note: Sparsity patterns specified are: $S$(HF2D13) $= \begin{bmatrix}
    1 & 0  & 1 & 0 \\ 0 & 1 & 0 & 1
\end{bmatrix},$ $ S$(WEC2) $= \begin{bmatrix}
    1 & 0 & 1 & 0 \\  0 &  1 & 0 &  1 \\ 1 & 0& 0 & 1
\end{bmatrix},$ $ S$(PSM) $=\begin{bmatrix}
    1 & 0 &1\\ 0 & 1 & 0
\end{bmatrix}, $ $S$(CDP) $=I.$ }
\end{table}
 We compare $SR_{la}$ and $SR_{sla}$ with the actual $SR$ values (computed using gradient based method in \cite{KATEWA2020108685}) in Table \ref{Tb:Tbl_SR_vals}. We denote the computation time (measured in seconds) for computation of $SR$, $SR_{la}$ and $SR_{sla}$ as $\tau, \tau_{la}$ and $\tau_{sla}$, respectively.
We observe that the approximation based algorithms work well in approximating the actual $SR$. Also, the SLA-based algorithm performs better than LA-based algorithm for the majority of systems. 
However, the former has a larger computation time since it is iterative in nature (c.f. Remark \ref{rem:Comp_LA_SLA}).
\subsection{System Design Problem}
In this subsection, we present numerical results for the system design problems \textbf{SD}$_{la}$ and \textbf{SD}$_{sla}$ given in \eqref{eq:eqn_ND_la_final} and \eqref{eq:eqn_ND_sla}. We use the MATLAB function \textit{fmincon} to solve these problems. In each iteration of \textit{fmincon}, we use the closed form expressions of $SR_{la}$ and $SR_{sla}$.

\begin{figure}[thpb]
\begin{subfigure}{\columnwidth}
    \centering
    \includegraphics[width=\columnwidth]{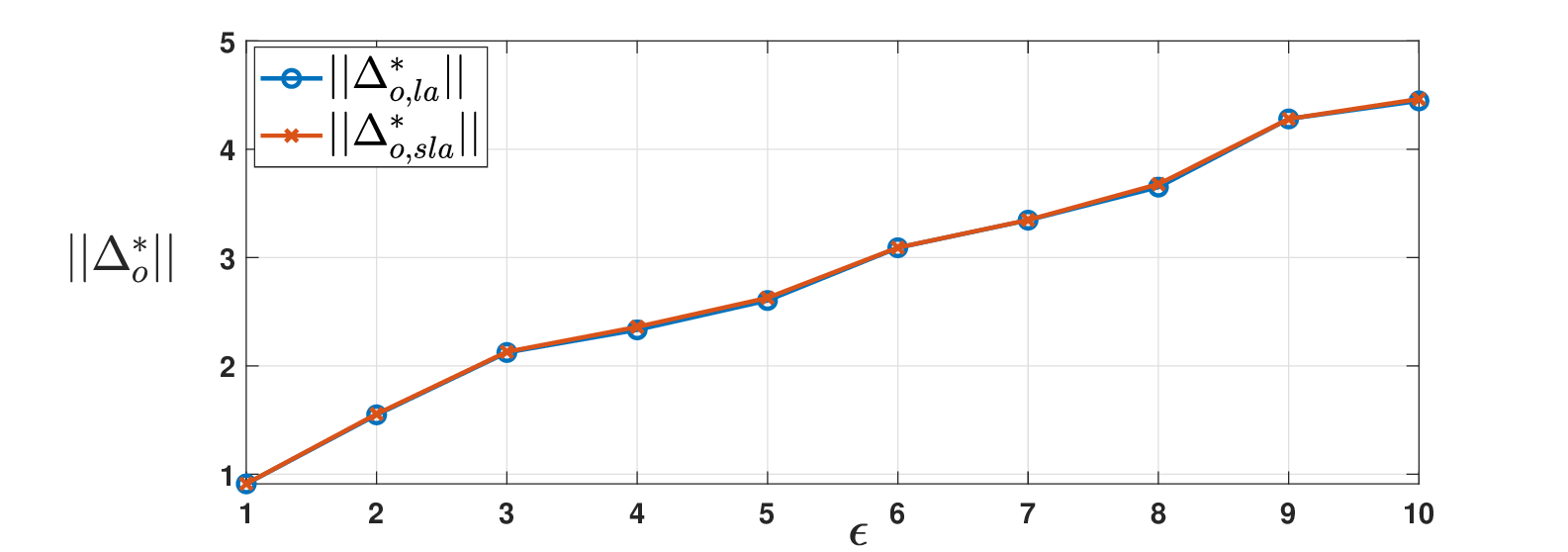}
    \caption{}
    \label{fg:sd_normal_deltao}
\end{subfigure}
\begin{subfigure}{\columnwidth}
    \centering
    \includegraphics[width=\columnwidth]{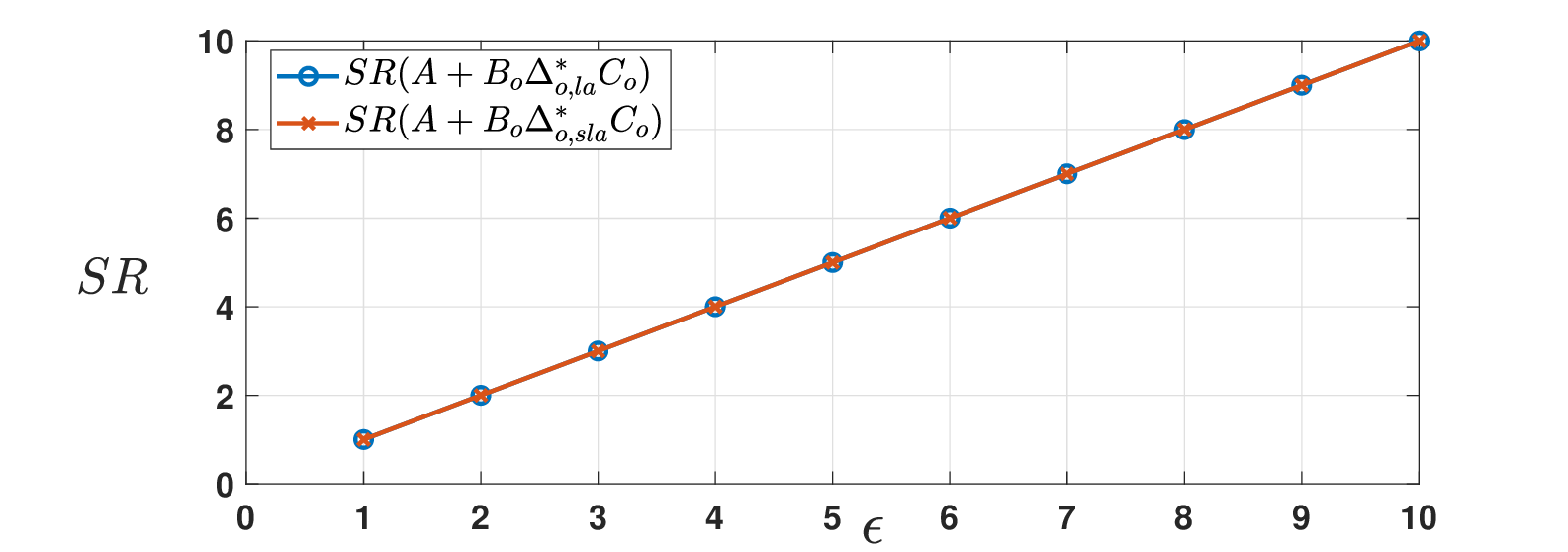}
    \caption{}
    \label{fg:sd_normal_sr}
\end{subfigure}      
\caption{Variation of (a) $\lVert \Delta_o ^{*} \rVert$,  and  (b) $SR$ as a function of $\epsilon$ for Case I.}
\label{fig:sd_normal}
\end{figure}

We begin by considering Case I and Case II presented in the previous subsection. We pick $B_o, C_o, S_o$ same as $B, C, S$ and solve \textbf{SD}$_{la}$ and \textbf{SD}$_{sla}$ for different values of the desired stability radius $\epsilon$. We denote the optimal solutions of \textbf{SD}$_{la}$ and \textbf{SD}$_{sla}$ as $\Delta_{o,la}^{*}$ and $\Delta_{o,sla}^{*}$, respectively. Figure \ref{fg:sd_normal_deltao} for Case I shows that the optimal perturbation norms $\lVert \Delta_{o,la}^{*}\rVert$ and $\lVert \Delta_{o,sla}^{*}\rVert$ are very close for all values of $\epsilon$. This is again due to the fact that $A$ is normal. Further, as $\epsilon$ increases, the optimal norms increase. This is because larger-norm perturbations to the system are required to achieve a larger value of $SR$. Further, Figure \ref{fg:sd_normal_sr} verifies that the actual $SR$ of the optimally perturbed system indeed matches with the desired $SR$ value $\epsilon$. 

\begin{figure}[thpb]
\begin{subfigure}{\columnwidth}
    \centering
    \includegraphics[width=\columnwidth]{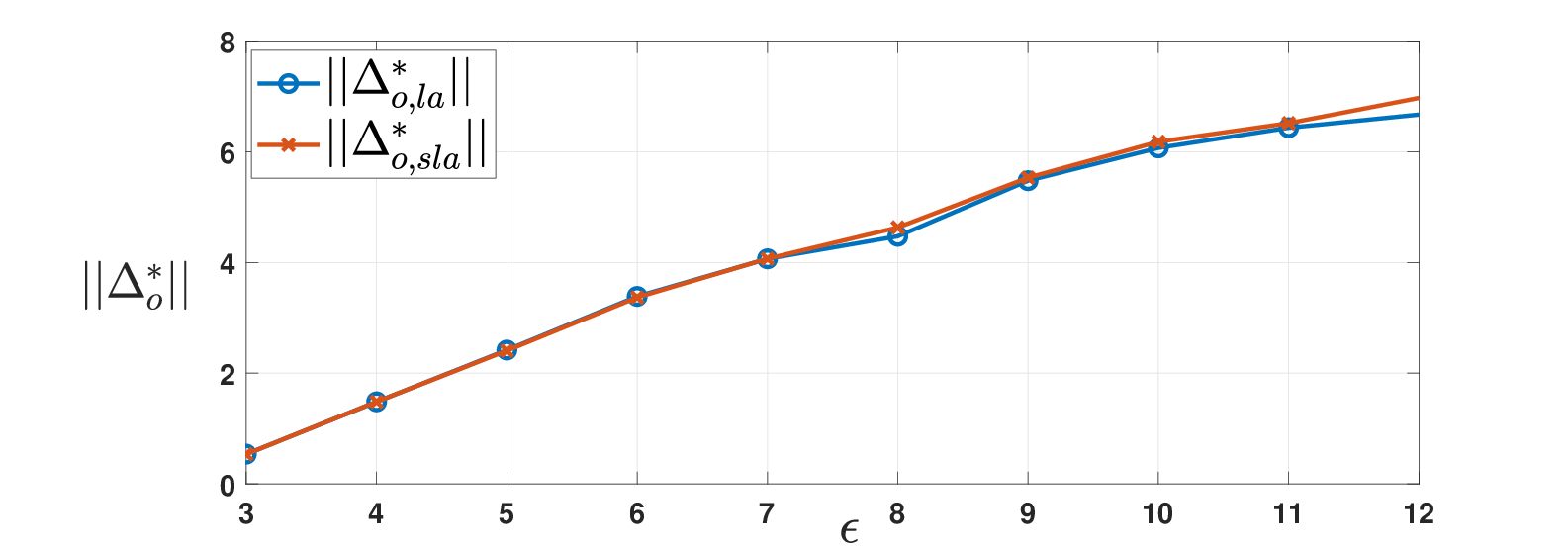} 
    \caption{}
    \label{fg:fig_sd_nonnorma_del_o}
\end{subfigure}
\begin{subfigure}{\columnwidth}
    \centering
    \includegraphics[width=\columnwidth]{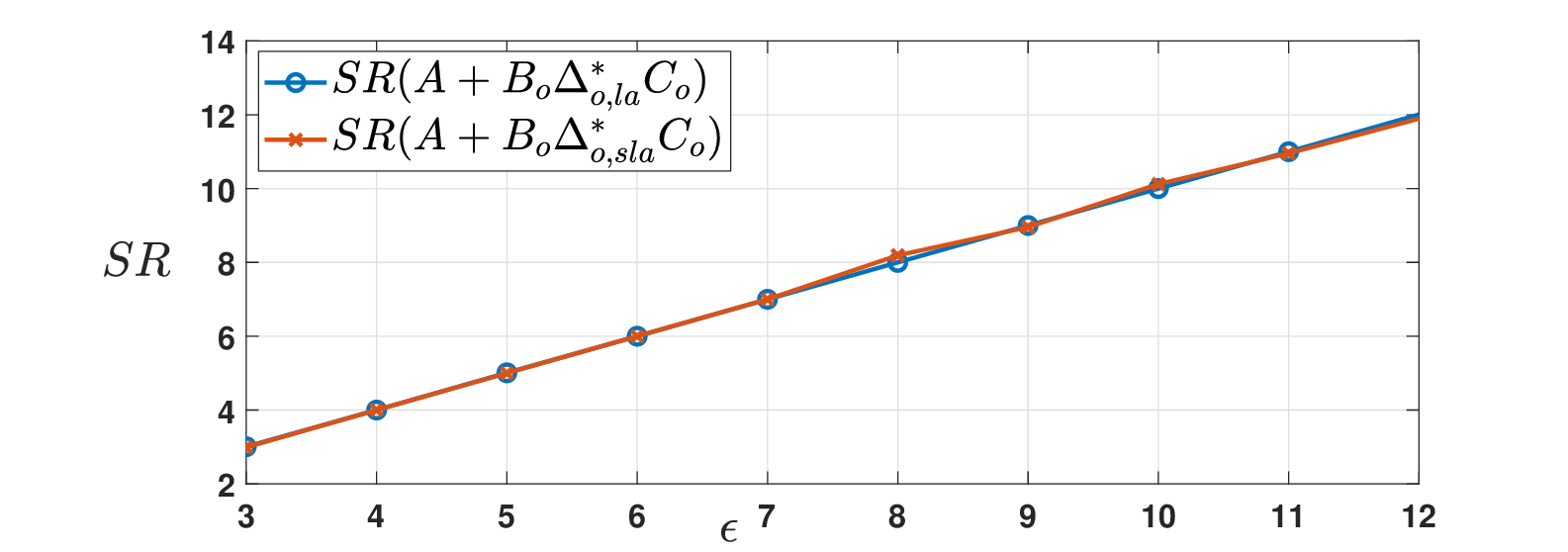}
    \caption{}
    \label{fg:fig_sd_nonnorma_sr}
\end{subfigure}
        
\caption{Variation of (a) $\lVert \Delta_o ^{*} \rVert$,  and  (b) $SR$ as a function of $\epsilon$ for Case II.}
\label{fig:sd_non_normal}
\end{figure}

The corresponding plots for Case II are presented in Figure \ref{fig:sd_non_normal}. In contrast to Case I, we observe some difference between the optimal solutions $ \Delta_{o,la}^{*}$ and $\Delta_{o,sla}^{*}$ for large values of $\epsilon$ in Figure \ref{fg:fig_sd_nonnorma_del_o}. Again, this is due to the fact that $A$ is non-normal in this case. In Figure \ref{fg:fig_sd_nonnorma_sr}, we observe that the modified system achieves the desired value of $SR$. 
Next, we consider the \textbf{SD}$_{la}$ and \textbf{SD}$_{sla}$ problems for the test problems given in Table \ref{Tb:Tbl_SD_vals}. For each system, we pick $B_o,C_o,S_o$ same as $B,C,S$ and set $\epsilon = 1.2 SR(A)$ (that is, we aim to increase the $SR$ by 20\%). 
Table \ref{Tb:Tbl_SD_vals} provides the norms of the optimal solutions. We denote the computation time (measured in seconds) for \textbf{SD}$_{la}$ and \textbf{SD}$_{sla}$ as $\tau_{o,la}$ and $\tau_{o,sla}$, respectively. 
We observe that $\lVert \Delta_{o,la}^{*} \rVert$ and $\lVert \Delta_{o,sla}^{*} \rVert$ are close for majority of problems and  $\lVert \Delta_{o,la}^{*} \rVert \geq \lVert \Delta_{o,sla}^{*} \rVert$ for all the cases. Further,  $\tau_{o,la} < \tau_{o,sla}$ for each problem as computation of $SR$ in \textbf{SD}$_{sla}$ involves successive-linear approximations.

These results demonstrate that our approximated $SR$ methods are useful for designing sparse and non-sparse systems with improved stability properties.

\begin{table}[h]
\begin{center}
\caption{$||\Delta_o^{*}||$ for different test problems}
\begin{tabular}{|c|c|c|c|c|}
\hline
 Test Problem & $ \lVert \Delta_{o,la}^{*}  \rVert$ &  $\lVert \Delta_{o,sla}^{*} \rVert$ & $\tau_{o,la}$ & $\tau_{o,sla}$ \\
 \hline
  Case I \cite{KATEWA2020108685} & 0.045 & 0.043  & 0.689 & 21.939\\
 \hline
 Case II \cite{KATEWA2020108685} & 0.4147 & 0.3631  & 10.23 &23.025 \\
 \hline
 HF2D12 & 0.1657 & 0.1649  & 20.74 & 22.1\\
 \hline
  HF2D13 (S) & 0.0502 & 0.0467 & 0.6105& 2.0420\\
 \hline
 TG1 &  0.0577 & 0.0520  &1.591 &3.579\\
 \hline
 AGS &0.1413 & 0.1141  & 5.472 &31.27\\
 \hline
 WEC2 (S) & 0.1015 & 0.1008 & 0.0426 & 1.5758\\
 \hline
 WEC3 & 0.5866   & 0.5721 & 12.31& 77.60\\
 \hline
 BDT1 & 0.0033 & 0.0030 & 12.309& 75.289 \\
 \hline
 MFP & 1.8646 & 1.5145  & 1.689& 6.014\\
 \hline
 UWV & 0.0944 & 0.0942 & 0.5410 & 4.141 \\
 \hline
 EB1 & 0.0036  & 0.0034 & 0.4773&4.898\\
 \hline
 PSM (S) &  0.4488 & 0.5078 & 1.9464  & 21.0752\\
 \hline
 CDP (S) & 1.6542 & 1.5195 &3.5064 & 153.42 \\
 \hline
\end{tabular}\label{Tb:Tbl_SD_vals} 
\end{center}
\end{table}

\section{CONCLUSION} \label{sec:Section_Conc}
We propose approximated $SR$ formulations based on eigenvalue sensitivity via linear and successive-linear approximations. We study these problems with sparsity constraints on perturbations and derive closed form solutions. These results are used to develop an efficient framework to improve the stability radius. Future works include proposing a relaxed convex version of the design problem to solve it efficiently and easily. Also, we aim to extend our analysis for other system properties like controllability, observability, detectibility, etc., to design system with improved characteristics.









\bibliographystyle{ieeetr} 
\bibliography{Refs} 

\end{document}